\documentclass[pdftex,twocolumn,epjc3]{svjour3}          

\RequirePackage[T1]{fontenc}

\smartqed  

\RequirePackage{graphicx}
\RequirePackage{mathptmx}      
\RequirePackage{flushend}
\RequirePackage[numbers,sort&compress]{natbib}
\RequirePackage[colorlinks,citecolor=blue,urlcolor=blue,linkcolor=blue]{hyperref}
\usepackage[mathlines]{lineno}
\journalname{Eur. Phys. J. C}

\newcommand {\sNN}[1]{$\sqrt{s_{\rm NN}} = #1$}
\newcommand {\pt}{p_{\rm T}}

\begin{document}

\title{Evidence of Mass Ordering of Charm and Bottom Quark Energy Loss in Au+Au Collisions at RHIC}


\author{STAR Collaboration \thanksref{t1, addr1}
}

\thankstext[$\star$]{t1}{e-mail: star-publication@bnl.gov}

\institute{Brookhaven National Laboratory, PO Box 5000, Upton, NY 11973-5000, USA\label{addr1}
}

\date{\today}


\maketitle

\begin{abstract}
Partons traversing the strongly interacting medium produced in heavy-ion collisions are expected to lose energy depending on their color charge and mass. We measure the nuclear modification factors for charm- and bottom-decay electrons, defined as the ratio of yields, divided by the number of binary nucleon-nucleon collisions, in $\sqrt{s_{\rm NN}}$ = 200 GeV Au+Au collisions to $p$+$p$ collisions ($R_{\rm AA}$), or in central to peripheral Au+Au collisions ($R_{\rm CP}$). We find the bottom-decay electron $R_{\rm AA}$ and $R_{\rm CP}$ to be significantly higher than those of charm-decay electrons. Model calculations including mass-dependent parton energy loss in a strongly coupled medium are consistent with the measured data. These observations provide evidence of mass ordering of charm and bottom quark energy loss when traversing through the strongly coupled medium created in heavy-ion collisions.
\end{abstract}

\section{Introduction}
Ultra-relativistic heavy-ion collision experiments at the Relativistic Heavy-Ion Collider (RHIC) and Large Hadron Collider (LHC) are unique in studying the strong interaction and underlying theory, Quantum Chromodynamics (QCD). Over the past decades, many experimental observations in these collisions have provided evidence that a novel QCD state of matter is created composed of de-confined quarks and gluons: the Quark-Gluon Plasma (QGP)~\cite{ADAMS2005102,BRAHMS:white:paper,PHENIX:white:paper,PHOBOS:white:paper,Bass_1999}. Heavy quarks, \textit{i.e.}, charm and bottom quarks, in heavy-ion collisions have emerged as essential probes of the QGP because they have a rest mass much larger than the expected QGP temperature. This restricts their production to the initial hard parton scatterings in the collision, and therefore they can carry information about the entire QGP evolution~\cite{PhysRevC.51.2177,PhysRevLett.95.122001,Rapp:2009my,Dong:2019byy}.

Heavy quark energy loss in the produced medium is expected to proceed via (quasi-)elastic scatterings with the medium constituents and induced gluon radiation~\cite{DOKSHITZER2001199, ABIR2012183}. Multiple elastic scatterings with the medium constituents also lead to a Brownian-like motion of low $p_{\rm T}$ heavy quarks. This transfers the collective motion of the expanding plasma to the heavy quark leading to large anisotropic flow. QCD predicts that heavy quarks lose less energy than light quarks due to the so-called "dead cone" effect~\cite{DOKSHITZER2001199, ABIR2012183}, \textit{i.e.}, gluon radiation is suppressed for heavy quarks at angles smaller than $\theta_{c} \sim M/E$ where $M$ and $E$ are the quark mass and its energy, and that gluons lose more energy than light quarks due to their larger color factor. Therefore, parton energy loss ($\Delta E$) in the QGP is expected to follow a hierarchy ordered by parton color charge and mass, \textit{i.e.}, $\Delta E(g)>\Delta E(u,d,s)>\Delta E(c)>\Delta E(b)$ with $g$, $u$, $d$, $s$, $c$, and $b$ denoting gluons, up, down, strange, charm, and bottom quarks, respectively~\cite{DOKSHITZER2001199, ABIR2012183}. Useful quantities to study parton energy loss are the nuclear modification factors, $R_{\rm AA}$ and $R_{\rm CP}$. $R_{\rm AA}$ is defined as the particle yield in heavy-ion collisions divided by the respective yield in $p$+$p$ collisions, scaled by the average number of binary nucleon-nucleon collisions ($N_{\rm coll.}$) in heavy-ion collisions:
\begin{center}
\begin{equation}
R_{\rm AA} =\frac{1}{N_{\rm coll.}}\times \frac{\mathrm{d}N^2_{\rm AA}/(\mathrm{d}\pt \mathrm{d}y)}{\mathrm{d}N^2_{\rm pp}/(\mathrm{d}\pt \mathrm{d}y)}.
\end{equation}
\end{center}
$R_{\rm CP}$ is defined as the ratio of the yield in head-on heavy-ion (central) collisions to the yield in collisions with small nuclear geometric overlap (peripheral), scaled by a factor to account for the different $N_{\rm coll.}$ in each case:
\begin{equation}
R_{\rm CP} = \frac{\mathrm{d}N^2/(\mathrm{d}\pt \mathrm{d}y)}{N_{\rm coll.}}|_{\rm central}\times \frac{N_{\rm coll.}}{\mathrm{d}N^2/(\mathrm{d}\pt \mathrm{d}y)}|_{\rm peripheral}.
\end{equation}
An observation of $R_{\rm AA}$ or $R_{\rm CP}$ that is equal to unity for heavy flavor hadrons would indicate heavy-ion collisions are an incoherent superposition of individual nucleon-nucleon collisions. It is worth noting that measurements of $R_{\rm AA}$ and $R_{\rm CP}$ are affected by both the energy loss and the steepness of the spectrum shape. For given energy loss, the steeper the spectrum shape, the smaller the $R_{\rm AA}$ and $R_{\rm CP}$.

At $p_{\rm T}<2$\,GeV/$c$, a "flow hump" is seen in charm hadron $R_{\rm AA}$ due to the collective radial flow transferred to the charm quark while at $p_{\rm T}>$ 5\,GeV/$c$, the charm hadron $R_{\rm AA}$ is much less than unity and is approximately the same as charged hadrons at sufficiently high $p_{\rm T}$ ($>$ 8-10\,GeV/$c$)~\cite{PhysRevC.99.034908,Abelev2012,Adam2016,Acharya2018,2018474,ALICE:2021rxa}. The interpretation of this observation is complicated by the interplay of other effects not related to medium induced energy loss, for example differences in the fragmentation functions and $p_{\rm T}$ spectra~\cite{RAPP201821,OHFOV1,OHFOV2,OHFOV3,PhysRevLett.112.042302}. The $R_{\rm AA}$ of bottom and charm hadrons are observed at the LHC to be similar within uncertainties for $p_{\rm T}$ larger than 7 GeV/$c$~\cite{PhysRevLett.119.152301,2018474}. Measurements at the LHC of bottom-decayed $J/\psi$ and $D^{0}$ $R_{\rm AA}$ at $p_{\rm T}<$ 20 GeV/$c$ show a hint of energy loss mass ordering when compared to measurements of prompt $D^{0}$ $R_{\rm AA}$~\cite{Sirunyan2018,PhysRevLett.123.022001,ATLAS:2018hqe}. The ALICE and ATLAS collaborations report separate values for bottom- and charm-decayed lepton $R_{AA}$ in the $p_{\rm T}$ interval [3,8] GeV/$c$ that are consistent with the mass ordering of energy loss~\cite{ALICE:2016uid,ATLAS:2021xtw,2021136558}. Models including the mass dependence of parton energy loss have predicted significantly different values of nuclear modification factors for bottom and charm hadrons and their decay leptons in heavy-ion collisions in the $p_{\rm T}$ ranges probed at RHIC~\cite{PhysRevLett.100.192301,PhysRevC.92.024907,PhysRevC.78.014904,PhysRevC.90.034910}. Therefore, a comparison of charm and bottom hadron nuclear modification factors at RHIC is an excellent probe of the expected hierarchy of parton energy loss. The PHENIX experiment has reported a hint of a bottom- and charm- decayed electron $R_{\rm AA}$ separation in the $p_{\rm T}$ interval of [3,4] GeV/$c$~\cite{PhysRevC.93.034904, PHENIX:BOTTOM}.  

In this paper, we report the $R_{\rm AA}$ of electrons\footnote{Unless specified otherwise, electrons referred here include both electrons and positrons.} from semileptonic decays of open charm and bottom hadrons in Au+Au collisions at $\sqrt{s_{\rm NN}}$ = 200 GeV. Additionally, we report the double-ratios of bottom- to charm-decay electron $R_{\rm AA}$ and $R_{\rm CP}$. The Au+Au results presented here include the measurement of the inclusive heavy flavor hadron decayed electron (HFE) spectra and the fraction of bottom-decay electrons to the sum of bottom- and charm-decay electrons ${f_{b}^{AA}\equiv~N(\it{b}\rightarrow\it{e})/N(\it{b}+\it{c}\rightarrow\it{e})}$. 

\section{Experiment and data analysis}
In this analysis, inclusive electrons consist of bottom- and charm-decay electrons, misidentified hadrons, photonic electrons from gamma conversions ($\gamma \rightarrow e^+e^-$) and Dalitz decays of $\pi^0$ and $\eta$ mesons ($\pi^0/ \eta \rightarrow e^+e^-\gamma$), hadron decayed electrons from prompt quarkonia, light vector mesons and kaon semi-leptonic decays ($K_{\rm e3}$), and Drell-Yan contributions. The $f_{b}^{AA}$ can be obtained by topologically separating the different electron sources in the inclusive electron sample by utilizing the distance of closest approach of the electron track with respect to the primary vertex~\cite{PhysRevC.99.034908}. The misidentified hadrons, photonic electrons, hadron decayed electrons, and Drell-Yan contribution are subtracted from the inclusive electrons in order to obtain inclusive HFE.

The data were collected by the Solenoidal Tracker At RHIC (STAR) experiment in 2014 and 2016. Minimum bias (MB) events are selected by requiring a coincidence between the Vertex Position Detectors (VPD)~\cite{LLOPE2004252} just outside the beampipe on both ends of the STAR detector and covering the pseudorapidity ($\eta$) range of 4.24 $< |\eta| <$ 5.1. An additional trigger is used to enrich the sample of high--$p_{\rm T}$ electrons by selecting events with a single Barrel ElectroMagnetic Calorimeter (BEMC)~\cite{BEDDO2003725} tower above a transverse energy threshold $E_{\rm T}>$ 3.5 GeV (denoted $high~tower$ or HT). This data sample corresponds to an integrated luminosity of 0.2 and 1.0 nb$^{-1}$ in 2014 and 2016 data, respectively. For the inclusive HFE measurement, the HT triggered data from the 2014 RHIC Run is utilized.

The main detectors used in the data analysis are the Time Projection Chamber (TPC)~\cite{ANDERSON2003659}, Time-Of-Flight (TOF) detector~\cite{LLOPE2012S110}, BEMC and Heavy Flavor Tracker (HFT)~\cite{CONTIN201860}. All of TPC, TOF and BEMC cover full azimuth within pseudo-rapidity range of $|\eta| < 1$. The TPC, a gas-filled detector, provides the complete tracking of charged particles, and is used for momentum determination and particle identification (PID) via measuring ionization energy loss ($dE/dx$). The TOF employs Multi-gap Resistive Plate Chamber (MRPC) technology and is used for PID by measuring the flight time of charged particles. The BEMC, a lead-scintillator sampling calorimeter, can identify high-$\pt$ electrons via the momentum to energy deposition ratio ($p/E$). The HFT, a four-layer high-resolution silicon vertex detector, provides a good track pointing resolution ($\sim$46 $\mu$m for 750 MeV/$c$ kaons projected to the collision vertex) that enables the topological separation of charm- and bottom-decay electrons.

\subsection{Fraction of bottom hadron decayed electrons}
For the $f_{b}^{AA}$ measurement, the primary vertex is required to be within 6 cm of the center of the STAR detector to ensure uniform tracking at mid-rapidity. Additionally, the difference between the TPC and VPD vertex $z$ position is required to be less than 3 cm to reduce events with multiple collisions. After event selection there are about 1 billion and 1.1 billion MB events in the 2014 and 2016 data samples, respectively. Tracking of charged particles up to $|\eta| < 1$ is achieved using the TPC inside a 0.5\,T magnetic field. Tracks are selected by requiring a minimum of 20 hits, out of a maximum of 45, in the TPC, and the ratio of recorded over possible hits is greater than 0.52 to reduce fake combinations. To improve $dE/dx$ resolution, the number of TPC hits used for $dE/dx$ calculation is required to be at least 15. Additionally, a distance-of-closest approach (DCA) to the primary vertex of less than 1 cm is applied. Reconstructed tracks are then matched to hits in the HFT detector.

\begin{figure}
	\includegraphics[scale=0.385]{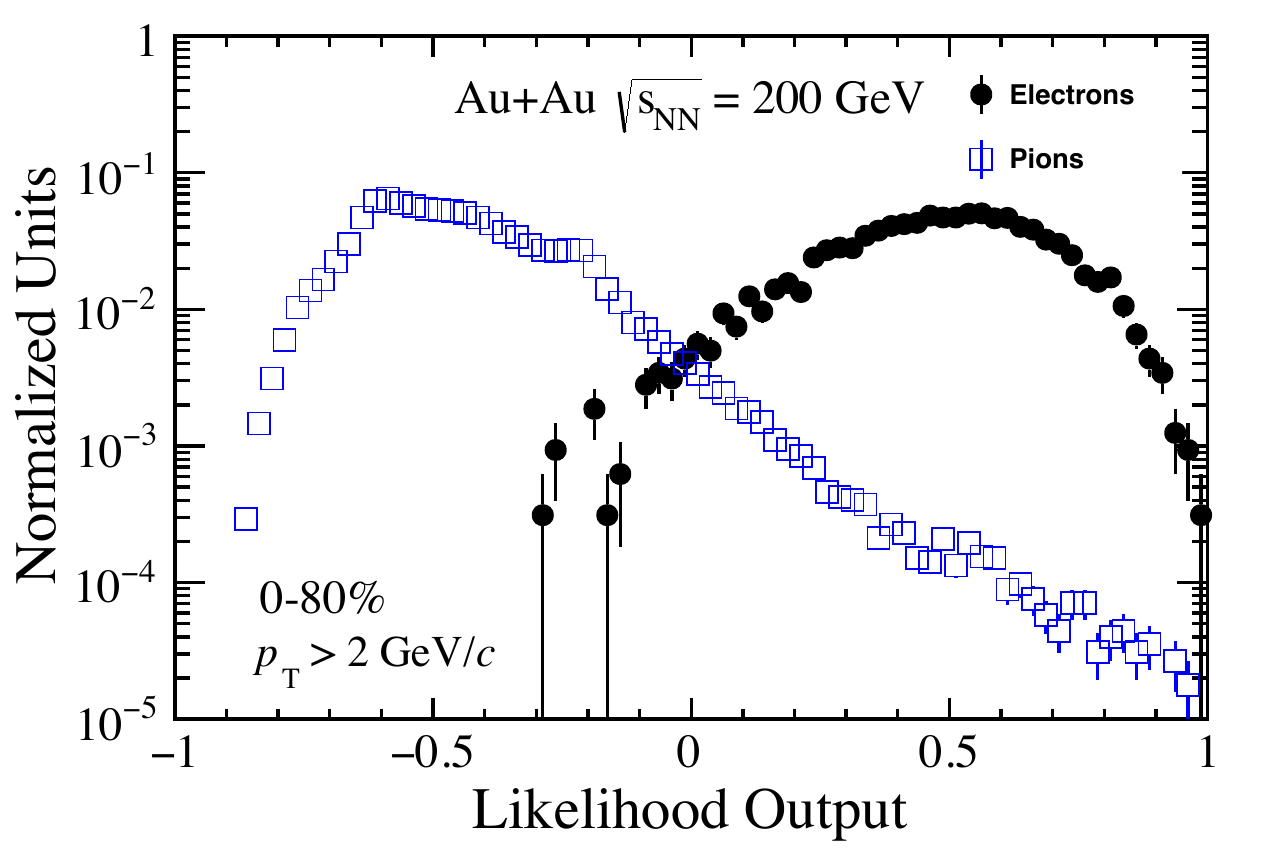}	
	\caption{
	Likelihood distribution for electrons (black circles) and pions (open blue squares) in 2014 data determined from the data-driven approach described in the text. \label{fig:Figure1S}}
\end{figure}

PID is conducted using a combination of the TPC, TOF, and the BEMC detectors. $dE/dx$ information in the TPC is used to select electrons by requiring $n\sigma_{e}$ to be between -1 and 3 for electrons, where $n\sigma_{e}$ is defined as $ln[(dE/dx)_{m}/(dE/dx)_{th}]/\sigma$ with $(dE/dx)_{m}$ and $(dE/dx)_{th}$ being the measured and expected $dE/dx$ for electrons, respectively, and $\sigma$ being the experimental resolution. For tracks with a $p_{\rm T}$ between 2 and 2.5 GeV/$c$ the TOF information is used to enhance the electron purity by requiring $|1/\beta-1|<0.025$, where $\beta$ is a track's velocity in units of speed of light. Finally, using the energy deposited in a single BEMC tower that is matched to the candidate electron track, we require the $p/E$ to be between 0.3 and 1.5. Since electrons tend to deposit all of their energy this value peaks around unity for electrons and is able to discriminate against hadrons, which tend to deposit a fraction of their energy producing a $p/E$ ratio larger than unity. Additional PID information is used in HT triggered data from the Shower Maximum Detectors (SMD) in the BEMC towers. The SMD is segmented in the $\eta$ and $\phi$ dimensions, and is able to distinguish between a broad electron and narrow hadron shower shape. We require the number of hits in both $\eta$ and $\phi$ to be greater than one. 

\begin{figure}
	\includegraphics[scale=0.415]{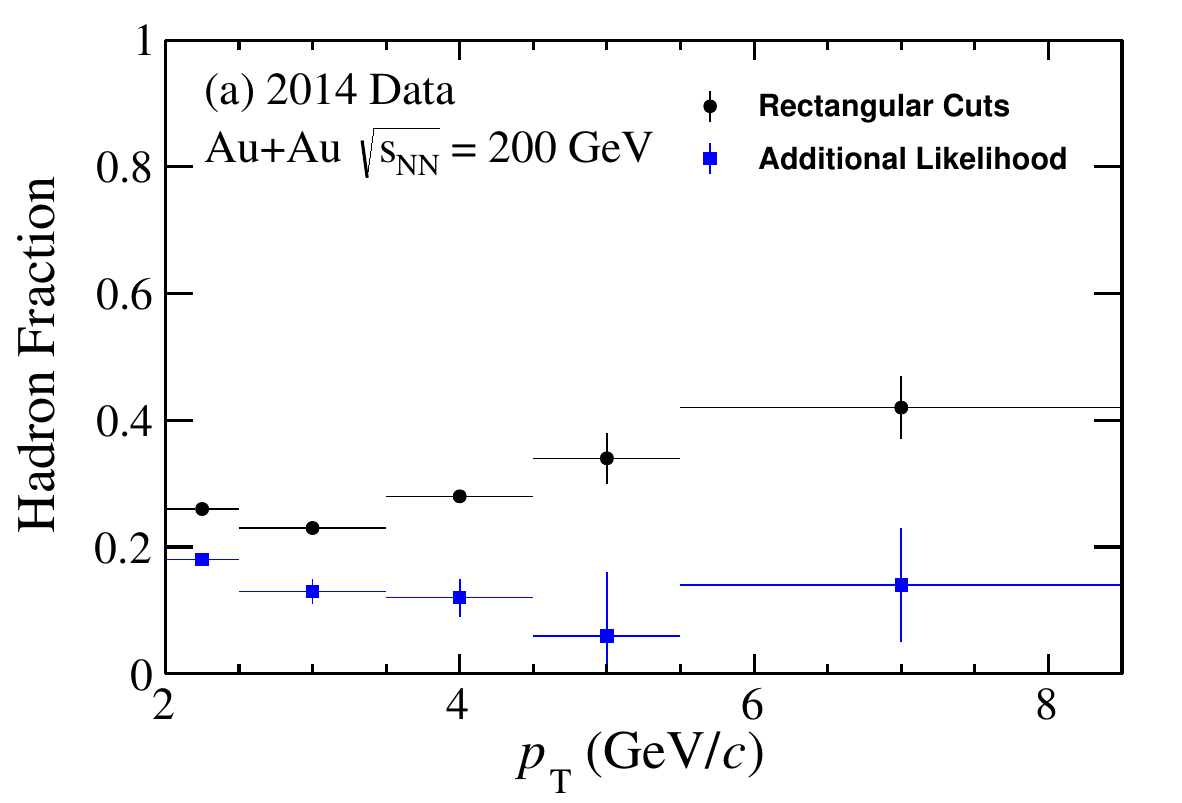}	
	\includegraphics[scale=0.415]{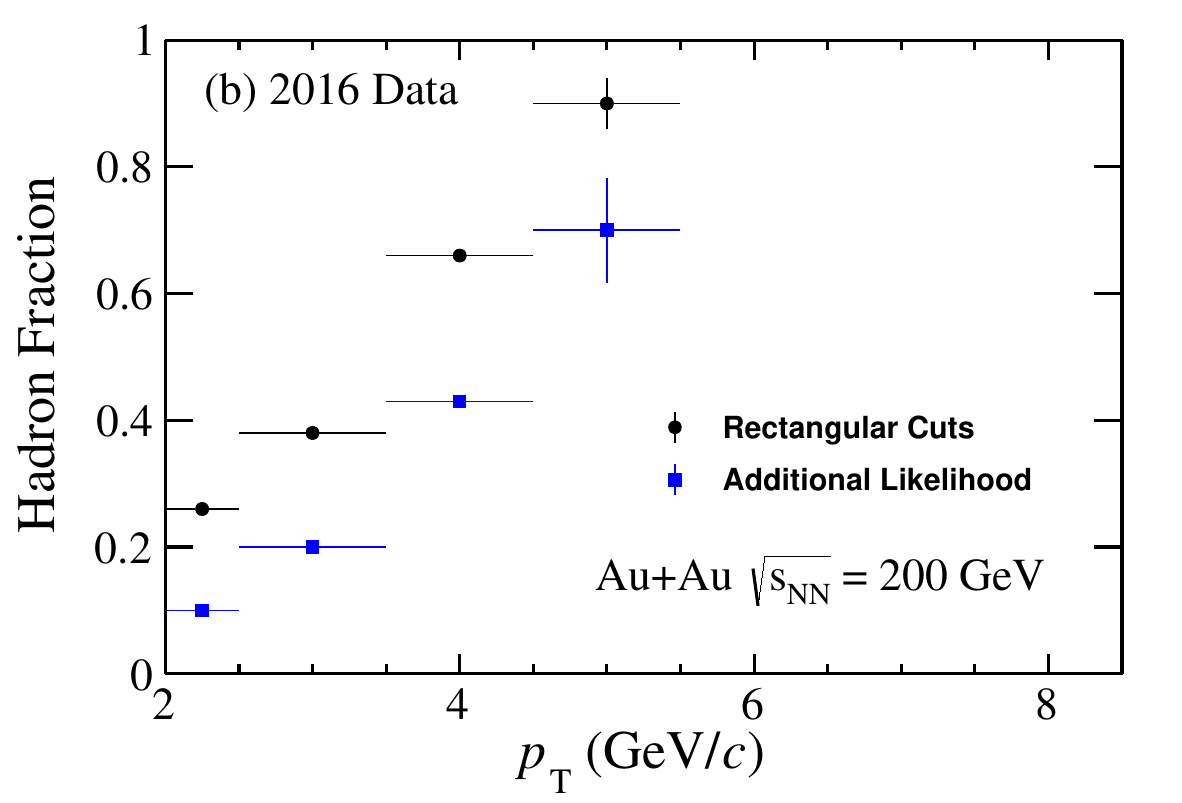}	
	\caption{Hadron contamination for rectangular particle identification (black circles) and with the additional likelihood selection (solid blue squares) in the remaining MB triggered sample described in the text for 2014 (a) and 2016 (b) data respectively. \label{fig:Figure2S}}
\end{figure}

A projective likelihood multivariate analysis (MVA) classifier, analogous to the TMVA method~\cite{Hocker:2007ht}, is used to further separate electrons from hadrons, which are predominantly pions, in the MB data sample using probability distribution functions (PDF) of electrons and pions in each PID sub-detector. Control samples of electrons and pions are used to determine the PDFs, and are constructed from photonic electrons and $K_{s}\rightarrow \pi^{+}\pi^{-}$ decays, respectively, using a tag-and-probe method. The likelihood is calculated as 
\begin{equation}\label{eq:like}
\mathcal{L} = \frac{\prod_{i}p^{e}_{i}}{\prod_{i}p^{e}_{i}+\prod_{i}p^{\pi}_{i}}, 
\end{equation}
where $p_{i}^{e/\pi}$ are the particle probabilities, and $i$ runs over all PID quantities. Equation~\ref{eq:like} is further transformed as $\mathcal{L}\rightarrow -1/15 \cdot ln(\mathcal{L}^{-1}-1)$ to have better discrimination between the signal and background peaks~\cite{Hocker:2007ht}. The variables used in the likelihood classifier are $n\sigma_{e}$, $p/E$, $1/\beta$ and the residuals in the $\phi$ and $z$ dimensions of the track projected onto the BEMC cluster center.
\begin{figure}
	\includegraphics[scale=0.415]{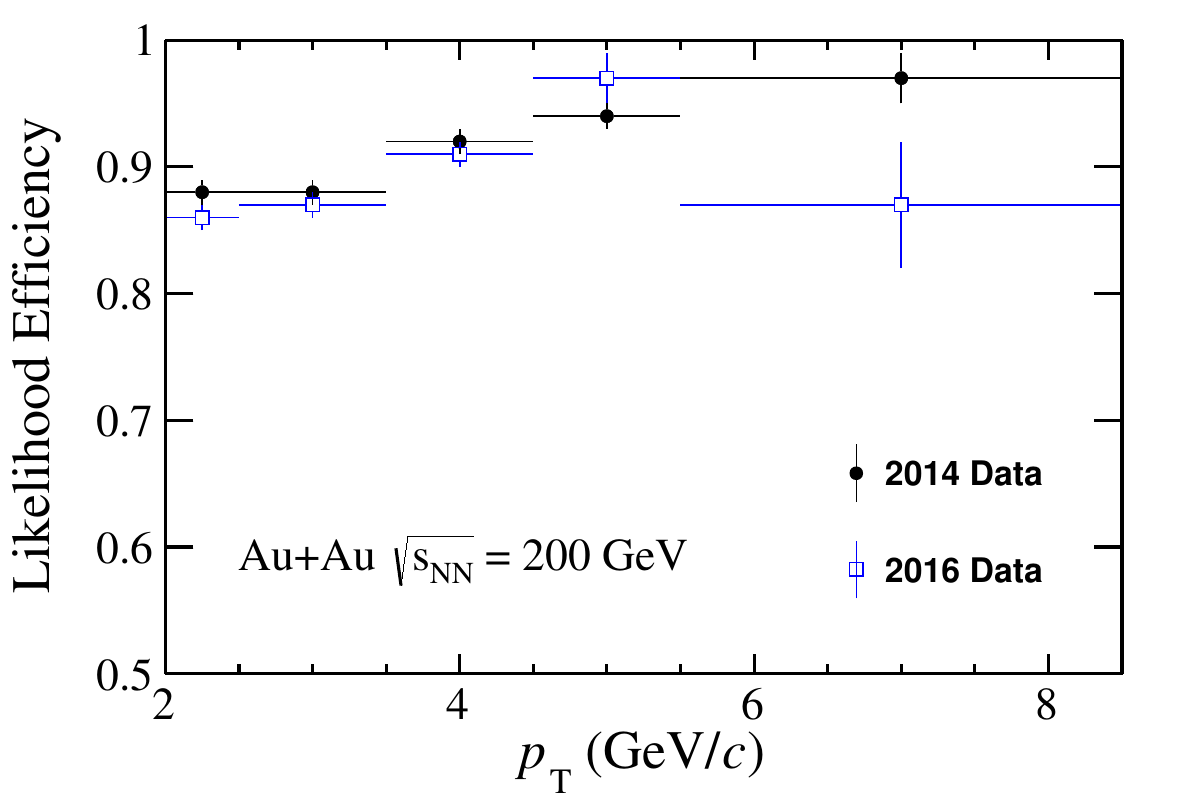}	
	\caption{The likelihood selection efficiency for 2014 (black circles) and 2016 (open blue squares) data. \label{fig:Figure3S}}
\end{figure}

The likelihood distribution for electrons and pions with no PID selections are shown in Fig.~\ref{fig:Figure1S} for 2014 data; similar distributions are observed for 2016 data. An optimization method is performed to maximize the electron purity without significantly reducing the signal efficiency, and corresponds to likelihood selections greater than 0.45 and 0.39 for 2014 and 2016 data, respectively. The hadron fractions before and after the likelihood selection are shown in Fig.~\ref{fig:Figure2S} for the remaining MB triggered sample, after taking out electron candidates that fire the HT trigger, in 2014 and 2016 data. It is observed that the likelihood selection provides a clear improvement over the standard rectangular selections. There are significant differences between the hadron fractions in 2014 and 2016 data. These are due to the STAR trigger configurations. In 2016, the high-tower triggers were utilized during most of the data taking and had significant overlap with the minimum bias triggers. Since the high-towers were designed to trigger on electrons in the $p_{\rm T}$ range used for this measurement, and additional particle identification from the shower maximum detectors could then be utilized, most electrons in minimum bias events are re-classified into the high-tower triggers resulting in a lower electron purity for the remaining MB triggered sample. In contrast, for 2014 data this was not the case, and the high-tower trigger sample represents a small fraction of the total data set. The efficiencies for the likelihood selections after standard PID selections, determined in the electron control sample, are shown in Fig.~\ref{fig:Figure3S} for 2014 and 2016 data, and are greater than 87\% in the $p_{\rm T}$ range used in this measurement. Therefore, the additional likelihood selection significantly reduces the hadron background without compromising the statistical precision of the signal electron sample. For 2014 and 2016 HT samples, the electron purity varies from 83\% to 92\% and 85\% to 95\% at 3.5 $<p_{\rm T}<$ 8.5 GeV/$c$, respectively.
\begin{figure}
			\includegraphics[scale=0.415]{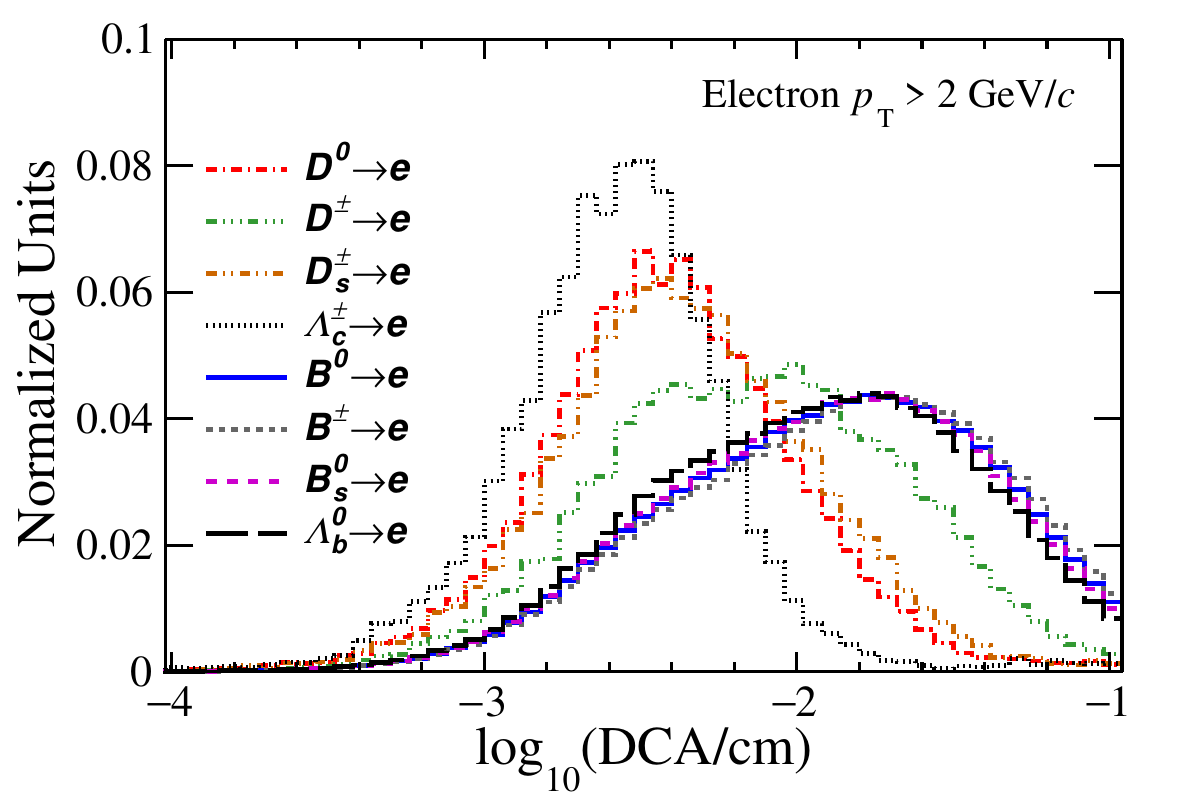}	
		\caption{Decay-electron ${\rm log}_{10}$(DCA/cm) distributions for different heavy flavor hadron species with an electron $p_{\rm T}>$~2.0\,GeV/$c$.    \label{fig:Figure6S}}
\end{figure}

We additionally reject electron candidates for which we find an oppositely charged electron in the event that produces a di-electron invariant mass lower than 0.15 GeV/$c^{2}$ to reduce photonic electron backgrounds. After electron identification requirements, the electron purity is greater than 80\% across all measured electron $p_{\rm T}$. Electron candidates are defined as tracks that pass all the above criteria.

\begin{figure}
    \includegraphics[scale=0.415]{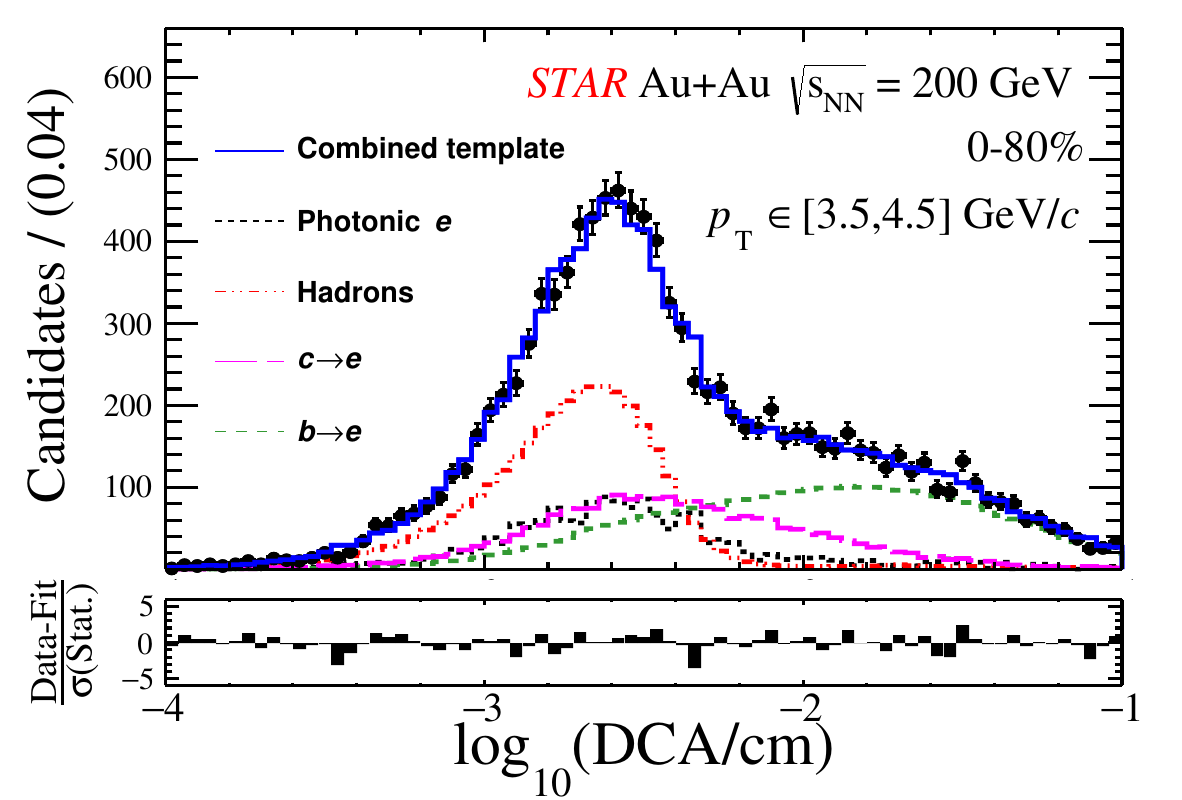}	
	\caption{
	Fit to the ${\rm log}_{10}$(DCA/cm) of candidate electrons with $p_{\rm T}\in$~[3.5,4.5]\,GeV/$c$ in 2014 data. The solid blue line shows the full template fit, and the various other lines show the individual components. The bottom panel shows the residual distribution of the template fit scaled by the statistical uncertainties. 
	\label{fig:Figure1}}
\end{figure}

We measure $f_{b}^{AA}$ in intervals of electron $p_{\rm T}$ by performing a four-component-template likelihood fit to the ${\rm log}_{10}$(DCA/cm) distribution of candidate electrons, where the DCA is defined as the 3D distance-of-closest approach of the track to the primary vertex. The hadron templates are taken from a sample of tracks selected with pion PID. The templates for residual photonic electrons are determined by embedding simulated detector hits of the decays of $\pi^{0}$, $\eta$, and photons in real data, and applying the same reconstruction and selection as data. The charm- and bottom-decay electron templates are constructed using the data-driven fast simulation technique described in Ref.~\cite{PhysRevC.99.034908}. All abundant ground states are included in the simulation. The initial charm hadron $p_{\rm T}$ spectra are taken from the measured $D^{0}$ spectra in Ref.~\cite{PhysRevC.99.034908}, and the relative hadron fractions are from available data for $D^{\pm}_{S}/D^{0}$~\cite{PhysRevLett.127.092301} and $\Lambda^{\pm}_{c}/D^{0}$~\cite{PhysRevLett.124.172301} or PYTHIA~\cite{SJOSTRAND2008852} for $D^{\pm}/D^{0}$. The PYTHIA calculation for $D^{\pm}/D^{0}$ is consistent with the ALICE measurement in Pb+Pb collisions~\cite{Acharya2018}. The bottom hadron spectra are taken from Fixed Order plus Next-to-Leading Logarithms (FONLL) calculations~\cite{Cacciari:2012ny,Cacciari:2015fta}, and we assume equal proportions of $B^{0}$ and $B^{\pm}$. The relative $B_{s}^{0}$ and $\Lambda_{b}^{0}$ fractions are taken from Ref.~\cite{PhysRevD.100.031102}. We include the contributions from $b\rightarrow c\rightarrow e$ decays in the bottom-decay electron templates. Due to the long and nearly identical bottom hadron lifetimes, we are not significantly sensitive to the bottom relative fractions. The decay-electron ${\rm log}_{10}$(DCA/cm) distributions for all heavy flavor hadron species that are considered in the fast simulation are shown in Fig.~\ref{fig:Figure6S} for an electron $p_{\rm T}\in$~[2.0,8.5]\,GeV/$c$. Potential backgrounds from Drell-Yan and prompt quarkonia are absorbed in the hadron template, as they produce electrons that point to the primary vertex. $K_{\rm e3}$ have DCA values outside the fit range considered. The only constrained normalization in the fit is the residual photonic electron template, which is estimated using a similar procedure as in Ref.~\cite{pphfe} to calculate the residual photonic electron yield. This yield relative to the candidate electron sample ranges from 25\% to 15\% from low to high $p_{\rm T}$, respectively. An example fit to 2014 data using the described templates is shown in Fig.~\ref{fig:Figure1}. 

The systematic uncertainties on the measured $f_{b}^{AA}$ fractions are: 1) the uncertainty on the simulated photonic electron cocktail composition is estimated by varying the relative proportions of electrons from photon conversions and light-meson decays by $\sim$50\%, and ranges from 8\% at low $p_{\rm T}$ to less than 1\% above 3.5\,GeV/$c$; 2) the uncertainty on the residual photonic electron template normalization is estimated by allowing it to vary by an absolute 5\%, and is about 3\%; 3) the charm hadron $p_{\rm T}$ spectra uncertainty is taken from the measured $D^{0}$ spectra, and is about 3\%; 4) the $D^{\pm}$/$D^{0}$ ratio uncertainty is 16\% from the differences in PYTHIA 6 and 8, and is 1--3\%; 5) the $\Lambda_{c}^{\pm}$/$D^{0}$ ratio is varied using the different models shown in Ref.~\cite{PhysRevLett.124.172301}, and the corresponding uncertainty is less than 1\%; 6) the uncertainty on the electron identification is estimated by tightening the selection, and is 2--3\% ; 7) the bottom hadron $p_{\rm T}$ spectrum uncertainty is estimated by applying both $b\rightarrow e$ and $c\rightarrow e$ $p_{\rm T}$ dependent $R_{\rm AA}$ calculated in the Duke model~\cite{PhysRevC.92.024907} described below, and is found to produce a maximum relative deviation of 2.5\% that is assigned across all electron $p_{\rm T}$ intervals.   
\begin{figure}
			\includegraphics[scale=0.285]{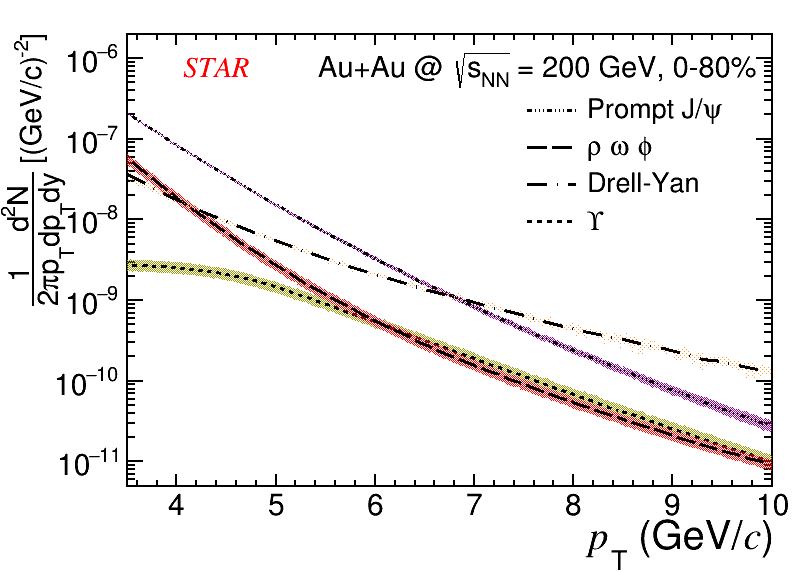}	
		\caption{Invariant yield of the electrons from decays of prompt $J/\psi$ (dot-dashed line), $\Upsilon$ (dotted line), Drell-Yan (long dashdotted line) and light vector mesons (long dashed line) in 0-80\% Au+Au collisions at $\sqrt{s_{\rm NN}}$ = 200 GeV. The bands represent systematic uncertainties.   \label{fig:Figure4S}}
\end{figure} 
\subsection{Inclusive heavy flavor hadron decayed electrons}
The invariant yield of inclusive HFE is measured using the same method as in Ref.~\cite{pphfe}. Compared to the $f_{b}^{AA}$ analysis, the notable differences are: 1) the primary vertex is required to be within 30 cm of the center of the STAR detector; 2) HFT hits are not included in track reconstruction; 3) maximum track DCA is 1.5 cm; and 4) $n\sigma_{e}$ is chosen from -1.5 to 3.0. The inclusive electron yield is first corrected for the mis-identified hadron contamination, which is a 4\% contribution at low $p_{\rm T}$ and 19\% at high $p_{\rm T}$. The photonic electron background is then subtracted using a data-driven method where low-mass electron pairs with opposite charges are reconstructed in data and efficiency-corrected to estimate the photonic electron yield, in which maximum di-electron mass is 0.24 GeV/$c^{2}$ and the minimum partner electron $p_{\rm T}$ is 0.3 GeV/$c$. The background-subtracted electron sample is then corrected for the tracking, PID, and trigger efficiencies which are calculated using the embedding technique. The total efficiency is 5\% at low $p_{\rm T}$ and 20\% at high $p_{\rm T}$. 
\begin{figure}
			\includegraphics[scale=0.28]{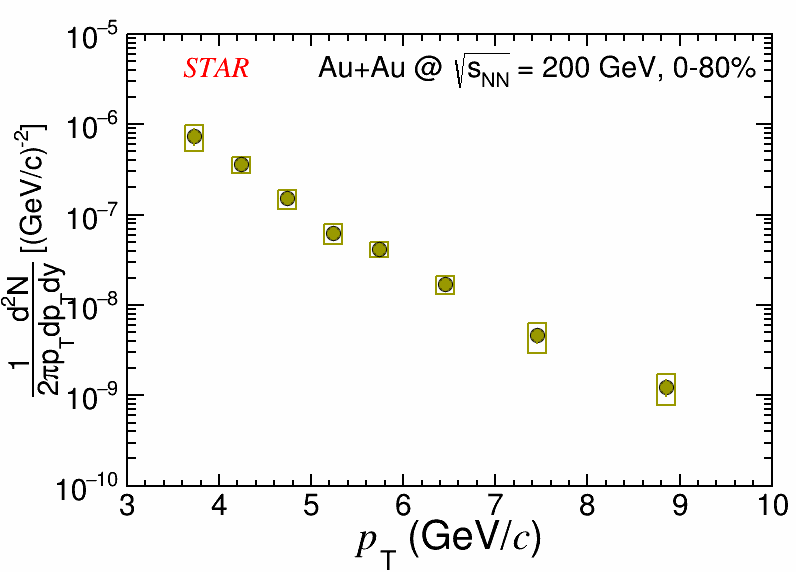}	
		\caption{The $\rm HFE$ invariant yield as a function of $\pt$ in 0-80\% Au+Au collisions at \sNN{200} GeV. The error bars and the boxes represent statistical and systematic uncertainties, respectively.    \label{fig:Figure5S}}
\end{figure}

We finally subtract the electron contributions from quarkonia, vector mesons, and Drell-Yan. The prompt $J/\psi$ decay background is subtracted using inclusive data~\cite{2019134917} with corrections to bottom hadron-decayed $J/\psi$ using FONLL with Color Evaporation Model (CEM) calculations~\cite{ref:fcem1,ref:fcem2}. EvtGen~\cite{LANGE2001152} is used to model the decay kinematics of quarkonium. Drell-Yan and Upsilon contributions are subtracted using $N_{\rm coll.}$-scaled PYTHIA and EvtGen, respectively, with the former taking no account of the nuclear and shadowing effects and the latter also incorporating the suppression model from~\cite{PhysRevD.97.016017}. The light-meson decays are estimated using $m_{\rm T}$-scaling of the $\pi^{0}$ data~\cite{PhysRevLett.97.152301, PhysRevLett.101.232301, PhysRevC.87.034911}. PYTHIA and EvtGen are used to model the electron decay channel of $\rho$, and $\omega$ and $\phi$, respectively. The $K_{\rm e3}$ also have a contribution to HFE yield, however, STAR simulation studies have shown that the $K_{\rm e3}$ contribution is less than 2\% at $p_{\rm T}>$ 3 GeV/$c$ in Au+Au collisions at $\sqrt{s_{\rm NN}}$ = 200 GeV~\cite{ref:ke3} and thus can be neglected. The obtained invariant yields of these contributions in the 0-80\% centrality interval of Au+Au collisions are shown in Fig.~\ref{fig:Figure4S}. These contributions amount to a $\sim$20\% reduction to the electron yield in the measured $p_{\rm T}$ region.

The systematic uncertainties considered in the inclusive heavy flavor electron production measurement, which in general increase with increasing $p_{\rm T}$, are the following. The electron reconstruction efficiency is evaluated by: 1) varying the required number of hits in the TPC for track reconstruction and $dE/dx$ from 20 and 15, to 25 and 18; 2) varying the maximum track DCA from 1.5 cm to 1.0 cm; and 3) varying the $p/E$ ratio in the BEMC between 0.6$<p/E<$1.5 and 0.3$<p/E<$1.8. The trigger efficiency uncertainty is estimated by varying the ADC trigger threshold in simulation by 3.5\%. The electron purity uncertainty and the efficiency of the electron identification selection are estimated by varying the mean and width parameters from the Gaussian fit to the pure electron $n\sigma_{e}$ distribution within one standard deviation from their central values. In addition, the latter includes an uncertainty from varying the $n\sigma_{e}$ selections from -1.5$<n\sigma_{e}<$3.0 to -1.0$<n\sigma_{e}<$3.0. The uncertainty on the photonic electron yield comes from the partner electron finding efficiency. The partner finding efficiency uncertainty is determined by varying the di-electron mass window from 0.24 to 0.15 GeV/$c^{2}$, and by varying the minimum partner electron $p_{\rm T}$ from 0.3 to 0.2 GeV/$c$. The uncertainties from the spectra of $\pi^{0}$ and $\eta$ mesons and their branching ratios, and from the partner electron tracking efficiency are also folded into the photonic electron uncertainty. The uncertainty from the quarkonia and vector meson subtraction comes from the uncertainty on the measured spectra, from the uncertainties on the FONLL+CEM calculations used to correct the inclusive $J/\psi$ spectra for non-prompt $J/\psi$, and from the uncertainty of the model calculation in Ref.~\cite{PhysRevD.97.016017} used to correct the Upsilon spectra. The uncertainty on the Drell-Yan contribution is evaluated in the same way as in Ref.~\cite{PhysRevD.99.072003}, taking into account the uncertainty of $N_{\rm coll.}$.

The obtained invariant yield of inclusive HFE ($\frac{e^++e^-}{2}$) in 0-80\% centrality of Au+Au collisions is shown in Fig.~\ref{fig:Figure5S}. 

\begin{figure}
		\includegraphics[scale=0.415]{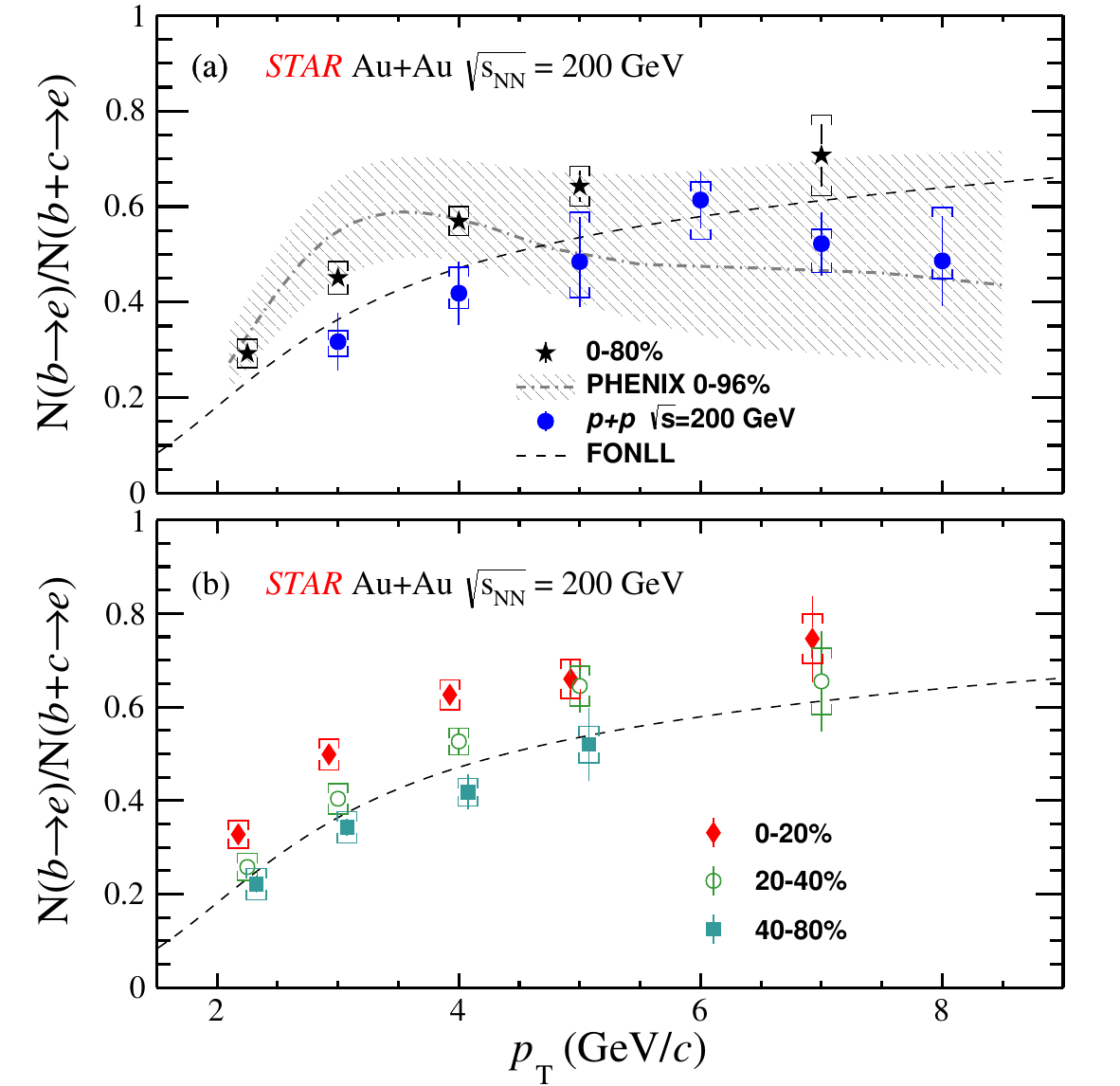}	
		\caption{The measured fraction of bottom hadron decayed electrons, $f_{b}^{AA}$, in intervals of electron $p_{\rm T}$ in Au+Au collisions at $\sqrt{s_{\rm NN}}=$ 200 GeV. (a) $f_{b}^{AA}$ in 0--80\% Au+Au collisions compared with the PHENIX measurements~\cite{PhysRevC.93.034904}, and the corresponding fraction $f_{b}^{pp}$ in $p$+$p$ collisions at $\sqrt{s}=$ 200 GeV~\cite{PhysRevLett.105.202301}. (b) $f_{b}^{AA}$ in three different centrality intervals. The error bars show statistical uncertainties, and the brackets show the systematic ones. The dashed line shows the central value of the FONLL prediction~\cite{Cacciari:2012ny,Cacciari:2015fta}. Data points are plotted along the $x$-axis at their respective bin centers, except 0--20\% and 40--80\% centrality classes, which are offset by 75 MeV/$c$ for clarity. \label{fig:Figure2}}
\end{figure}

\section{Results}
From the likelihood fit to the ${\rm log}_{10}$(DCA/cm) distribution of candidate electrons shown in Fig.~\ref{fig:Figure1}, the measured values for $f_{b}^{AA}$ from the combined 2014 and 2016 data samples are shown in Fig.~\ref{fig:Figure2}(a) along with the PHENIX measurements ~\cite{PhysRevC.93.034904} for the 0-80\% centrality class, and are compared to $p$+$p$ data~\cite{PhysRevLett.105.202301} and FONLL predictions. These STAR results in Au+Au collisions are consistent within uncertainties with published PHENIX measurements~\cite{PhysRevC.93.034904}. The $f_{b}^{AA}$ fractions in the 0--20\%, 20--40\%, and 40--80\% centrality regions are shown in Fig.~\ref{fig:Figure2}(b). Centrality is defined using the charged particle multiplicity at midrapidity~\cite{PhysRevC.79.034909}, and is related to the impact parameter of the colliding nuclei. The 0--20\% class includes nuclear collisions with the greatest spacial overlap, while 40--80\% denotes peripherally colliding nuclei. A clear centrality dependence is shown for $p_{\rm T}<$~4.5\,GeV/$c$, with significantly enhanced $b\rightarrow e$ fractions in 0--80\% and 0--20\% collisions compared to $p$+$p$ data and FONLL predictions. The results in 40--80\% collisions are in good agreement with the $p$+$p$ data and FONLL predictions.
\begin{figure}
			\includegraphics[scale=0.415]{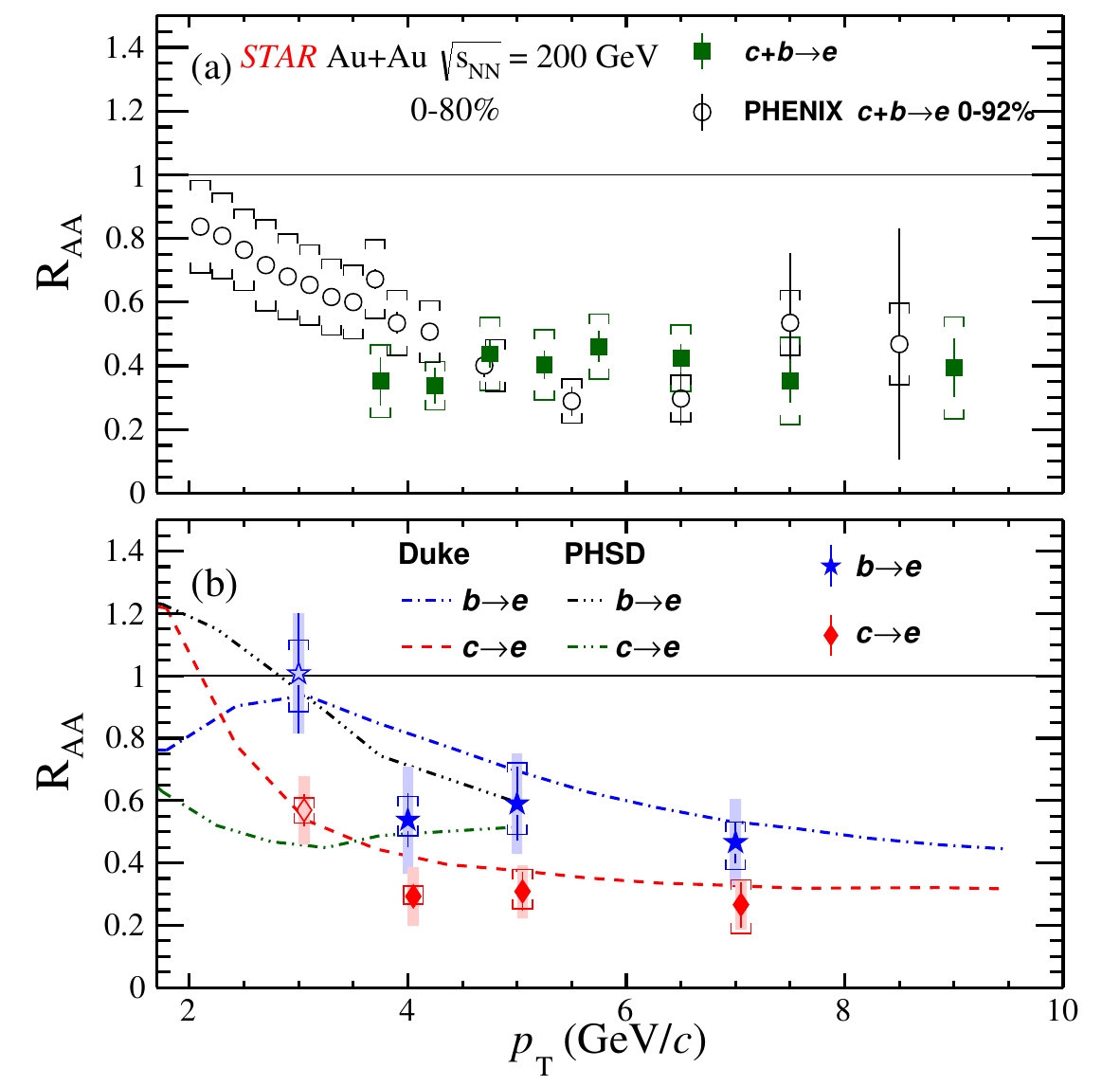}	
		\caption{(a) The inclusive heavy flavor electron $R_{\rm AA}^{incl.}$ (green squares) and the measurement from PHENIX (open circles)~\cite{PhysRevC.84.044905} in intervals of electron $p_{\rm T}$ in Au+Au collisions at $\sqrt{s_{\rm NN}}=$ 200 GeV. (b) The measured $R_{\rm AA}$ for bottom- (blue stars) and charm-decay (red diamonds) electrons in intervals of electron $p_{\rm T}$ in Au+Au collisions at $\sqrt{s_{\rm NN}}=$ 200 GeV. The open markers indicate values calculated using the PHENIX $R_{\rm AA}^{incl.}$. In all panels the error bars show the statistical uncertainties, and the brackets show the systematic ones. The shaded boxes on the data in (b) show the uncertainty due to $R_{\rm AA}^{incl.}$. Additional 8($\oplus$8)\% and 8.1\% uncertainties from the $N_{\rm coll.}$ calculations ($p$+$p$ luminosity) for the STAR and PHENIX $R_{\rm AA}$, respectively, are not shown. The Duke~\cite{PhysRevC.92.024907} and PHSD~\cite{Cassing:2008sv,Cassing:2009vt} models are shown as the various lines in (b). Data points are plotted along the $x$-axis at their respective bin centers, except the charm-decay electron data in (b), which are offset by 50 MeV/$c$ for clarity. \label{fig:Figure3}}
\end{figure}

The HFE nuclear modification factors, $R_{\rm AA}^{incl.}$, are calculated using the HFE production yields in Au+Au collisions shown in Fig.~\ref{fig:Figure5S} and $p$+$p$ reference data from Ref.~\cite{pphfe}, and are shown in Fig.~\ref{fig:Figure3}(a). The $R_{\rm AA}^{incl.}$ is compatible with the PHENIX measurement~\cite{PhysRevC.84.044905}. We decompose the charm- and bottom-decay electron $R_{\rm AA}$ using the measured fractions $f_{b}$ in Au+Au and $p$+$p$ collisions: $R_{\rm AA}^{b\rightarrow e} = f_{b}^{AA}/f_{b}^{pp} \times R_{\rm AA}^{incl.}$ and $R_{\rm AA}^{c\rightarrow e} = (1-f_{b}^{AA})/(1-f_{b}^{pp}) \times R_{\rm AA}^{incl.}$. For the $p_{\rm T}\in$~[2.5,3.5] GeV/$c$, we use the PHENIX inclusive heavy flavor electron measurement. The charm- and bottom-decay electron $R_{\rm AA}$ values are shown in Fig.~\ref{fig:Figure3}(b).

\begin{figure}
			\includegraphics[scale=0.415]{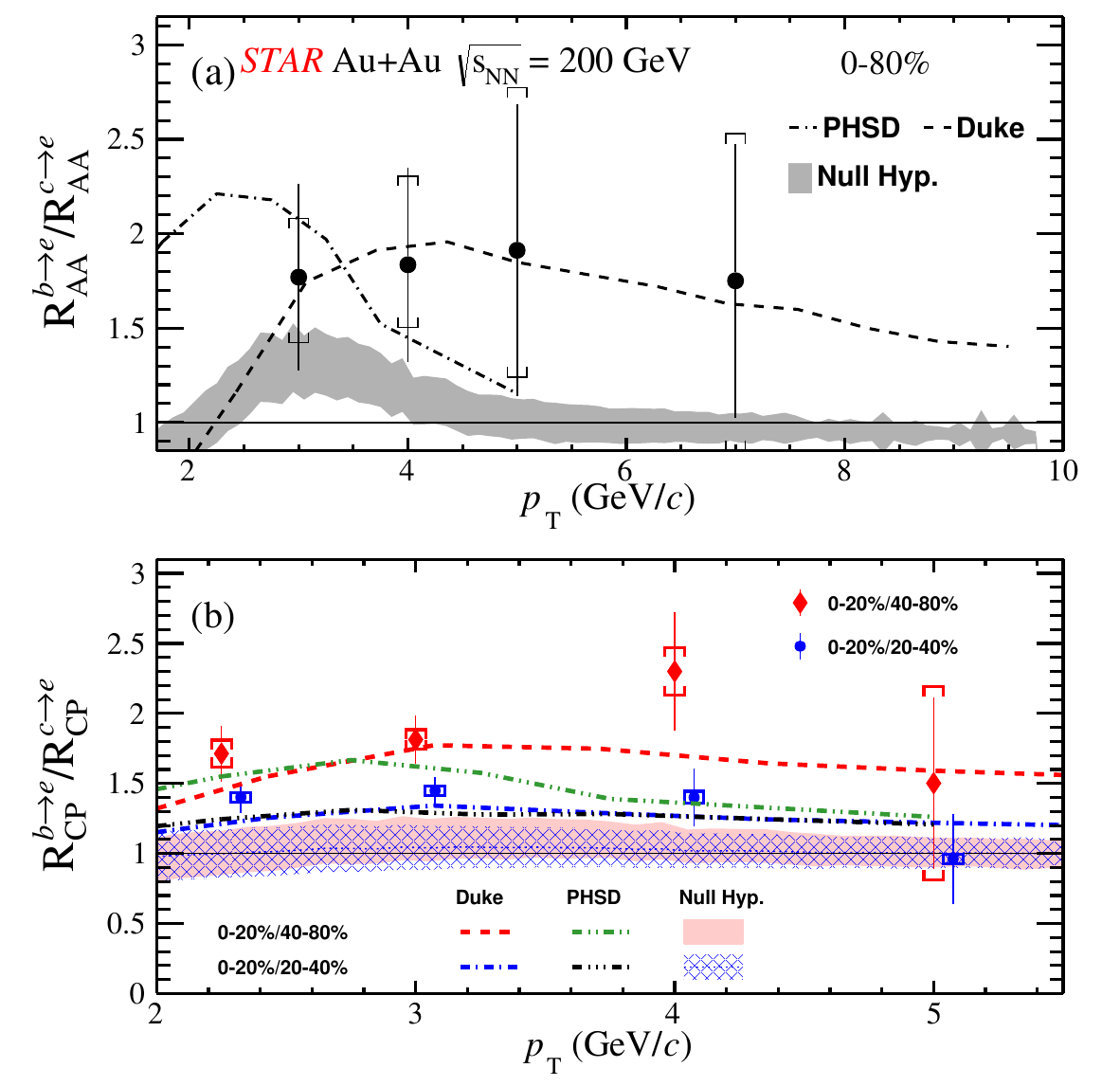}	
		\caption{(a) The $R_{\rm AA}$ ratio of bottom- to charm-decay electrons in intervals of electron $p_{\rm T}$ in Au+Au collisions at $\sqrt{s_{\rm NN}}=$ 200 GeV. (b) The $R_{\rm CP}$ ratios of bottom-decay electrons to that of charm-decay electrons in intervals of electron $p_{\rm T}$ in Au+Au collisions at $\sqrt{s_{\rm NN}}=$ 200 GeV. The red diamonds show the ratios of $R_{\rm CP}(0-20\%/40-80\%)$, and the blue circles show the ratios of $R_{\rm CP}(0-20\%/20-40\%)$. In all panels the error bars and the brackets show statistical and systematic uncertainties, respectively. The Duke~\cite{PhysRevC.92.024907} and PHSD~\cite{Cassing:2008sv,Cassing:2009vt} models are shown as the various lines. The null hypothesis calculations are shown as the shaded bands (see text for details). $R_{\rm AA}$ and $R_{\rm CP}(0-20\%/40-80\%)$ points are plotted along the $x$-axis at their respective bin centers and $R_{\rm CP}(0-20\%/20-40\%)$ is shifted by 75 MeV/$c$ for clarity.    \label{fig:Figure4}}
\end{figure}

The ratios of bottom- and charm-decay electron $R_{\rm AA}$ and $R_{\rm CP}$ are shown in Fig.~\ref{fig:Figure4}(a) and (b), respectively. Note that the ratios do not depend on $R_{\rm AA}^{incl.}$. The data show that bottom-decay electron $R_{\rm AA}$ compared to charm-decay electrons are systematically larger, with a central value about 80\% larger. The $R_{\rm CP}$ ratios for $R_{\rm CP}(0-20\%/40-80\%)$ and $R_{\rm CP}(0-20\%/20-40\%)$ show a more significant deviation from unity compared to $R_{\rm AA}$ because the systematic uncertainties largely cancel in the ratio. 

The kinematic shift of semileptonic decays, heavy quark production spectra, and the heavy flavor hadrochemistry may cause the decay-electron double ratios of $R_{\rm AA}$ and $R_{\rm CP}$ to differ from unity in the case that the bottom and charm hadron $R_{\rm AA}$/$R_{\rm CP}$ are the same. We incorporate these effects in a null hypothesis for the ratios of bottom- and charm-decay electron nuclear modification factors using simulated charm and bottom hadron decays with the same $p_{\rm T}$ dependent $R_{\rm AA}$ or $R_{\rm CP}$, which have initial spectra from perturbative QCD~\cite{Cacciari:2012ny,Cacciari:2015fta}. The abundances of $\Lambda_{c/b}$ are matched to data in Refs.~\cite{Acharya20182,PhysRevD.100.031102}. We then multiply the hadron spectra for both charm and bottom hadrons by the $p_{\rm T}$ dependent Duke model $R_{\rm AA}$ or $R_{\rm CP}$ values for $D$ mesons shown in ~\cite{PhysRevC.99.034908}, and propagate to the final state electrons. We then take the ratios of bottom- and charm-decay electron nuclear modification factors. In Fig.~\ref{fig:Figure4}, these are shown as the shaded bands labeled ``Null Hyp.". We performed systematic variations by changing the $D(B)_{s}$ and $\Lambda_{c(b)}$ fractions by 50\% in both $p$+$p$ and Au+Au, and the Duke model $R_{\rm AA}$ and $R_{\rm CP}$ values by a relative 25\% in each case. The most significant effect comes from the charm baryon fractions in $p$+$p$ and Au+Au collisions.

We perform a null hypothesis t-test, including data and null hypothesis model uncertainties, and find the bottom-to-charm $R_{\rm AA}$ ratios are consistent with the null hypothesis in the $p_{\rm T}\in$~[2.5,4.5]\,GeV/$c$ range. The bottom-to-charm $R_{\rm CP}$(0--20\%/40--80\%) and $R_{\rm CP}$(0--20\%/20--40\%) ratios in the $p_{\rm T}\in$~[2.0,4.5]\,GeV/$c$ range reject the null hypothesis at 4.2 and 3.3 standard deviations, respectively. In all t-tests, bin-by-bin correlations of systematic uncertainties in data are included. The PHENIX experiment also measured the $R_{\rm AA}$ of bottom- and charm-decay electrons for different centrality bins in Au+Au collisions at $\sqrt{s_{\rm NN}} = 200\,\,\rm GeV$~\cite{PhysRevC.93.034904, PHENIX:BOTTOM}, and found that charm-decay electrons are more suppressed than bottom-decay electrons with a significance of at least one standard deviation for 0–40\% central collisions. Compared to PHENIX $R_{\rm AA}$ results, our $R_{\rm CP}$ measurements with a significance larger than 3 standard deviations have a much improved precision.

We compare our results of charm- and bottom-decay electron nuclear modification factors to theoretical models describing the heavy-quark dynamics in the de-confined medium in Figs.~\ref{fig:Figure3} and \ref{fig:Figure4}. The curves denoted ``PHSD" show the Parton-Hadron-String-Dynamics model~\cite{Cassing:2008sv,Cassing:2009vt} and the ``Duke" curves show a modified Langevin transport model~\cite{PhysRevC.92.024907}. Both models include heavy quark diffusion in the QGP medium, heavy quark hadronization through coalescence and fragmentation, and mass-dependent energy loss mechanisms. In our measured electron $p_{\rm T}$ region, the corresponding parent heavy flavor hadrons $\langle p_{\rm T}\rangle$ covers about 4 - 8 GeV/$c$, where the contribution from heavy quark collectivity to the measured $R_{\rm AA}/R_{\rm CP}$ is negligible~\cite{PhysRevC.99.034908}. Additionally, the modification in heavy flavor hadrochemistry in Au+Au collisions does not play a significant role to account for the observed $R_{\rm AA}/R_{\rm CP}$ suppression. This is because the coalescence contribution is expected to be small in the measured $p_{\rm T}$ region~\cite{PhysRevC.92.024907,Plumari2018}, and the contribution from heavy flavor strange mesons and heavy flavor baryons to the electrons is small~\cite{Si:2019esg}. We find that both models can reproduce the bottom- and charm-decay electron $R_{\rm AA}$ and ratios of $R_{\rm CP}$, suggesting the mass ordering of parton energy loss in the QGP medium.


\section{Summary}
We have measured $R_{\rm AA}$ of inclusive heavy flavor-decay electrons, and separately for bottom- and charm-decay electrons in the $p_{\rm T}$ range of 2 to 8.5 GeV/$c$ in $\sqrt{s_{\rm NN}}$ = 200 GeV Au+Au collisions. In addition, we have measured the double ratios of bottom- and charm-decay electron $R_{\rm AA}$ and $R_{\rm CP}$. We find the measured values of bottom-decay nuclear modification factors are systematically larger than those of charm-decay electrons after accounting for effects not related to parton energy loss. The significance of this observation, in the $p_{\rm T}$ range of 2 to 4.5 GeV/$c$, is 4.2 and 3.3 standard deviations for the double ratios of $R_{\rm CP}(0-20\%/40-80\%)$ and $R_{\rm CP}(0-20\%/20-40\%)$, respectively. Compared to the data, the Duke and PHSD models are compatible within experimental uncertainties. These observations represent a significant comparison of bottom and charm hadron energy loss in heavy-ion collisions at RHIC, and provide evidence of mass ordering of charm and bottom quark energy loss when traversing through the strongly coupled medium created in heavy-ion collisions.

\begin{acknowledgements}
We thank the RHIC Operations Group and RCF at BNL, the NERSC Center at LBNL, and the Open Science Grid consortium for providing resources and support.  This work was supported in part by the Office of Nuclear Physics within the U.S. DOE Office of Science, the U.S. National Science Foundation, National Natural Science Foundation of China, Chinese Academy of Science, the Ministry of Science and Technology of China and the Chinese Ministry of Education, the Higher Education Sprout Project by Ministry of Education at NCKU, the National Research Foundation of Korea, Czech Science Foundation and Ministry of Education, Youth and Sports of the Czech Republic, Hungarian National Research, Development and Innovation Office, New National Excellency Programme of the Hungarian Ministry of Human Capacities, Department of Atomic Energy and Department of Science and Technology of the Government of India, the National Science Centre and WUT ID-UB of Poland, the Ministry of Science, Education and Sports of the Republic of Croatia, German Bundesministerium f\"ur Bildung, Wissenschaft, Forschung and Technologie (BMBF), Helmholtz Association, Ministry of Education, Culture, Sports, Science, and Technology (MEXT) and Japan Society for the Promotion of Science (JSPS).
\end{acknowledgements}

\bibliographystyle{spphys}
\bibliography{svjourn3-epjc}

\bfseries\raggedright\sffamily{
M.~S.~Abdallah$^{4}$,
B.~E.~Aboona$^{53}$,
J.~Adam$^{15}$,
L.~Adamczyk$^{2}$,
J.~R.~Adams$^{38}$,
J.~K.~Adkins$^{30}$,
I.~Aggarwal$^{40}$,
M.~M.~Aggarwal$^{40}$,
Z.~Ahammed$^{59}$,
D.~M.~Anderson$^{53}$
% ...
}



\normalfont\small\itshape
\begin{list}{}{%
}

\item{$^{1}$Abilene Christian University, Abilene, Texas   79699}
\item{$^{2}$AGH University of Science and Technology, FPACS, Cracow 30-059, Poland}
\item{$^{3}$Argonne National Laboratory, Argonne, Illinois 60439}
\item{$^{4}$American University of Cairo, New Cairo 11835, New Cairo, Egypt}
\item{$^{5}$Ball State University, Muncie, Indiana, 47306}
\item{$^{6}$Brookhaven National Laboratory, Upton, New York 11973}
\item{$^{7}$University of Calabria \& INFN-Cosenza, Italy}
\item{$^{8}$University of California, Berkeley, California 94720}
\item{$^{9}$University of California, Davis, California 95616}
\item{$^{10}$University of California, Los Angeles, California 90095}
\item{$^{11}$University of California, Riverside, California 92521}
\item{$^{12}$Central China Normal University, Wuhan, Hubei 430079 }
\item{$^{13}$University of Illinois at Chicago, Chicago, Illinois 60607}
\item{$^{14}$Creighton University, Omaha, Nebraska 68178}
\item{$^{15}$Czech Technical University in Prague, FNSPE, Prague 115 19, Czech Republic}
\item{$^{16}$Technische Universit\"at Darmstadt, Darmstadt 64289, Germany}
\item{$^{17}$ELTE E\"otv\"os Lor\'and University, Budapest, Hungary H-1117}
\item{$^{18}$Frankfurt Institute for Advanced Studies FIAS, Frankfurt 60438, Germany}
\item{$^{19}$Fudan University, Shanghai, 200433 }
\item{$^{20}$University of Heidelberg, Heidelberg 69120, Germany }
\item{$^{21}$University of Houston, Houston, Texas 77204}
\item{$^{22}$Huzhou University, Huzhou, Zhejiang  313000}
\item{$^{23}$Indian Institute of Science Education and Research (IISER), Berhampur 760010 , India}
\item{$^{24}$Indian Institute of Science Education and Research (IISER) Tirupati, Tirupati 517507, India}
\item{$^{25}$Indian Institute Technology, Patna, Bihar 801106, India}
\item{$^{26}$Indiana University, Bloomington, Indiana 47408}
\item{$^{27}$Institute of Modern Physics, Chinese Academy of Sciences, Lanzhou, Gansu 730000 }
\item{$^{28}$University of Jammu, Jammu 180001, India}
\item{$^{29}$Kent State University, Kent, Ohio 44242}
\item{$^{30}$University of Kentucky, Lexington, Kentucky 40506-0055}
\item{$^{31}$Lawrence Berkeley National Laboratory, Berkeley, California 94720}
\item{$^{32}$Lehigh University, Bethlehem, Pennsylvania 18015}
\item{$^{33}$Max-Planck-Institut f\"ur Physik, Munich 80805, Germany}
\item{$^{34}$Michigan State University, East Lansing, Michigan 48824}
\item{$^{35}$National Institute of Science Education and Research, HBNI, Jatni 752050, India}
\item{$^{36}$National Cheng Kung University, Tainan 70101 }
\item{$^{37}$Nuclear Physics Institute of the CAS, Rez 250 68, Czech Republic}
\item{$^{38}$Ohio State University, Columbus, Ohio 43210}
\item{$^{39}$Institute of Nuclear Physics PAN, Cracow 31-342, Poland}
\item{$^{40}$Panjab University, Chandigarh 160014, India}
\item{$^{41}$Purdue University, West Lafayette, Indiana 47907}
\item{$^{42}$Rice University, Houston, Texas 77251}
\item{$^{43}$Rutgers University, Piscataway, New Jersey 08854}
\item{$^{44}$Universidade de S\~ao Paulo, S\~ao Paulo, Brazil 05314-970}
\item{$^{45}$University of Science and Technology of China, Hefei, Anhui 230026}
\item{$^{46}$South China Normal University, Guangzhou, Guangdong 510631}
\item{$^{47}$Shandong University, Qingdao, Shandong 266237}
\item{$^{48}$Shanghai Institute of Applied Physics, Chinese Academy of Sciences, Shanghai 201800}
\item{$^{49}$Southern Connecticut State University, New Haven, Connecticut 06515}
\item{$^{50}$State University of New York, Stony Brook, New York 11794}
\item{$^{51}$Instituto de Alta Investigaci\'on, Universidad de Tarapac\'a, Arica 1000000, Chile}
\item{$^{52}$Temple University, Philadelphia, Pennsylvania 19122}
\item{$^{53}$Texas A\&M University, College Station, Texas 77843}
\item{$^{54}$University of Texas, Austin, Texas 78712}
\item{$^{55}$Tsinghua University, Beijing 100084}
\item{$^{56}$University of Tsukuba, Tsukuba, Ibaraki 305-8571, Japan}
\item{$^{57}$United States Naval Academy, Annapolis, Maryland 21402}
\item{$^{58}$Valparaiso University, Valparaiso, Indiana 46383}
\item{$^{59}$Variable Energy Cyclotron Centre, Kolkata 700064, India}
\item{$^{60}$Warsaw University of Technology, Warsaw 00-661, Poland}
\item{$^{61}$Wayne State University, Detroit, Michigan 48201}
\item{$^{62}$Yale University, New Haven, Connecticut 06520}
\end{list}

\end{document}